\documentclass[twocolumn,aps,pre,amsmath,amssymb,superscriptaddress,longbibliography]{revtex4-2}
\usepackage{graphicx}
\usepackage{bm}
\usepackage[colorlinks=true,linkcolor=blue,citecolor=blue,urlcolor=blue]{hyperref}
\begin{document}
\title{Temperature as a Dynamically Maintained Steady State:\\
Photonic Mechanisms, Maintenance Cost, and the Limits of the\\
Infinite-Reservoir Idealization}
\author{David Vaknin}
\email{vaknin@ameslab.gov}
\affiliation{Ames National Laboratory, Ames, Iowa 50011, USA}
\affiliation{Department of Physics and Astronomy, Iowa State University,
Ames, Iowa 50011, USA}
\date{\today}
\begin{abstract}
Classical thermodynamics treats temperature as a state variable
characterizing systems in equilibrium with idealized infinite reservoirs.
We argue that this framing, while computationally exact, obscures an
essential physical reality: any real open system at finite characteristic
energy $E_c = k_B T$ continuously exchanges energy with its environment
through thermal radiation; without compensating input, the system cools
toward the environmental temperature scale.
What thermodynamics calls ``thermal equilibrium'' is, for such open
systems, a dynamically sustained steady state maintained by ongoing
electromagnetic energy exchange.
We resolve the apparent contradiction with the purely mechanical Maxwell
velocity distribution: billiard-ball kinetics correctly describe the
\emph{shape} of the distribution at a given $E_c$, but cannot account
for how $E_c$ is established or maintained against radiative losses in
any real open system.
We further show that every finite thermal reservoir is itself maintained
by energy exchange at a larger scale, with the classical infinite reservoir
emerging as the large-capacity limit of a physical hierarchy rather
than a fundamental entity, and with electromagnetic radiation providing
a ubiquitous boundary channel through which temperature is stabilized
at each level.
\end{abstract}
\maketitle
\section{Introduction: The fiction of thermal equilibrium}
\label{sec:intro}
\subsection{The central claim}
Classical thermodynamics describes systems ``at temperature $T$'' as
though temperature were an intrinsic property like mass or charge.
The formalism is elegant, predictive, and spectacularly successful.
It is also incomplete in a fundamental way that has been systematically
overlooked.
\textbf{In real open systems, a constant temperature is physically
realized as a dynamically maintained condition, not a passive
static label.}
We emphasize immediately what this means and what it does not mean.
Formal statistical mechanics defines thermal equilibrium as a
stationary state with detailed balance. It does not claim that
microscopic processes cease in equilibrium: emission and absorption,
scattering, and fluctuations continue. Our claim is not against the
formal definition. It is that the \emph{physical realization} of
temperature in actual matter---composed of charged particles interacting
via the electromagnetic field---is physically realized through ongoing
energy exchange with the environment, with electromagnetic radiation
serving as a ubiquitous and often dominant channel. This is the
mechanism the formalism deliberately abstracts away. The common
pedagogical picture of ``equilibrium'' as a state of stasis, while
convenient, is physically incorrect for any real open system at
$E_c > 0$.
Any real open system at finite characteristic energy $E_c = k_B T$
continuously exchanges energy with its environment through thermal
radiation. Without compensating input, the system cools: $E_c$
decreases toward the temperature scale set by the environment.
Maintaining constant $E_c$ requires energy exchange with the
environment balancing radiative losses. What classical thermodynamics
calls ``thermal equilibrium'' is, in physical terms, a dynamically
sustained steady state.
This is not a subtle correction. It is the difference between
describing a waterfall as a static configuration of water molecules
and recognizing it as sustained flow. The distribution is steady, but
the constituents are in continuous exchange. Temperature characterizes
not a static state of matter but the flux of photons through it.
\subsection{Summary of contributions}
\textbf{(1) Steady state, not stasis.} We argue that the common framing
of ``thermal equilibrium'' as a static or passive state is physically
misleading for any real open system at $E_c > 0$. Such systems
continuously exchange energy with their environment through
electromagnetic radiation. Without compensating input the system cools;
maintaining constant $E_c$ requires ongoing energy exchange at the
system boundary. The correct physical description is \emph{steady
state}: the distribution is stationary, but its constituent excitations
are in continuous exchange (Sec.~\ref{sec:steadystate}).
\textbf{(2) Quantitative characterization of photon exchange.} We
derive that the mean photon energy in the Planck distribution is
$\langle h\nu \rangle \approx 2.701\,E_c$ (Sec.~\ref{sec:dynamic}).
This characterizes the typical energy scale of photons exchanged in
any radiatively coupled system maintaining temperature $T$.
\textbf{(3) The Maxwell distribution objection resolved.} A natural
objection is that the Maxwell velocity distribution of a gas follows
from purely mechanical collisions without photons. We show that
billiard-ball kinetics correctly give the \emph{shape} of the
distribution but cannot maintain $E_c$ over time in any real system
of charged particles subject to radiative losses. We further show
that proposed non-photonic mechanisms such as phonon-phonon scattering
are themselves electromagnetic at the foundational level
(Sec.~\ref{sec:maxwell}).
\textbf{(4) Hierarchy of finite reservoirs.} Every finite thermal bath
is maintained by energy exchange at a larger scale. The classical
infinite reservoir is not a physical entity but the large-capacity
limit of a hierarchy extending from individual samples to stellar
interiors (Sec.~\ref{sec:hierarchy}).
\textbf{(5) Dimensionless entropy.} We clarify the relation between
thermodynamic entropy $S = k_B \ln W$ and the dimensionless entropy
$\mathcal{S} = \ln W$, highlighting that $k_B$ serves to express
entropy in conventional SI units and to link microstate counts to
macroscopic observables (Sec.~\ref{sec:entropy}).
\subsection{Scope}
We focus on systems where electromagnetic coupling dominates energy
exchange: atoms, molecules, condensed phases, and plasmas. This
encompasses essentially all laboratory, terrestrial, and most
astrophysical systems. Systems thermalized by other interactions
(neutrinos in supernovae, gluons in quark-gluon plasma) require
analogous treatment with appropriate force carriers; the underlying
principle---that temperature is a dynamically maintained steady state
sustained by continuous exchange of the relevant bosonic field---applies
broadly. We note that even in condensed-matter systems where energy
is redistributed internally by phonons, electron-phonon coupling, or
spin exchange, the macroscopic temperature of the system is
ultimately set by energy exchange at its boundaries, with
electromagnetic radiation providing a ubiquitous and often dominant
channel for that boundary exchange (see Sec.~\ref{sec:maxwell}).

\subsection{The spirit of this work}

We emphasize that the physical observations assembled here are not,
individually, new. Spontaneous emission and its role in detailed
balance have been understood since Einstein~\cite{einstein1917}.
Stefan-Boltzmann radiation from any body above absolute zero is a
standard result~\cite{jackson1999}. The $H$-theorem and its domain
of applicability have been analyzed for over a century~\cite{boltzmann1872}.
The finite nature of real thermal reservoirs is well appreciated in
experimental low-temperature physics~\cite{pobell2007}.

What we offer is not new physics but a synthesis: when these
well-established facts are drawn together, they form a consistent
physical picture in which what statistical mechanics calls
``thermal equilibrium at temperature $T$'' is, in any real open
system, a dynamically maintained condition sustained by ongoing
electromagnetic energy exchange with the environment. Classical
thermodynamics and statistical mechanics are not wrong; they are
silent on mechanism by design. Our aim is to supply that mechanism
explicitly, making the physical picture consistent with what is
actually happening in the systems the formalism describes.

The misconception we address is not found in the formal literature.
It is found in the physical image that standard presentations
inadvertently convey: that a system ``at equilibrium'' is simply
sitting at a fixed temperature, with nothing required to keep it
there. That image is at odds with Einstein's $A$ coefficients,
with Stefan-Boltzmann emission, and with the experimental reality
that every thermostated system requires a power source. Our
contribution is to make this tension explicit and to show how it
resolves naturally once the maintaining mechanism is identified.

\section{Historical foundation: From Planck to quantum thermodynamics}
\label{sec:history}
The recognition that temperature has quantum foundations emerged from
the blackbody radiation problem. Classical physics predicted spectral
energy density diverging catastrophically at high frequencies---the
ultraviolet catastrophe~\cite{rayleigh1900,jeans1905}. Planck's 1900
resolution~\cite{planck1901} introduced energy quantization:
electromagnetic oscillators exchange energy only in discrete units
$E = h\nu$, yielding the correct distribution
\begin{equation}
u(\nu) = \frac{8\pi h\nu^3}{c^3}\frac{1}{e^{h\nu/k_BT}-1},
\label{eq:planck_hist}
\end{equation}
eliminating the divergence through exponential suppression.
The profound implication emerged gradually: thermal radiation---the
primary mechanism of energy exchange between matter and environment---is
fundamentally quantum mechanical. The existence of thermal distributions
at temperature $T$ necessarily entails photons obeying
Eq.~\eqref{eq:planck_hist}.
Einstein's 1917 analysis of radiative transitions~\cite{einstein1917}
demonstrated that equilibrium between matter and radiation requires
three processes: absorption, stimulated emission, and crucially,
\emph{spontaneous emission}---photon emission without external field.
The necessity of spontaneous emission showed that matter at finite
$E_c$ inevitably radiates. Without spontaneous emission, detailed
balance fails and the Planck distribution cannot be
maintained~\cite{milonni1994,scully1997}. Einstein's coefficients
ensure that equilibrium populations follow the Boltzmann distribution
when photon exchange occurs according to Planck statistics, establishing
temperature as a property of systems in \emph{continuous} photon
exchange rather than as a passive state variable.
Boltzmann's statistical mechanics~\cite{boltzmann1872,boltzmann1877}
and Gibbs' ensemble theory~\cite{gibbs1902} provided powerful frameworks
for equilibrium thermodynamics, but deliberately abstracted away from
microscopic processes. Gibbs introduced the canonical ensemble by
assuming systems exchange energy with a heat bath at fixed temperature,
without specifying what physically maintains that temperature. The
present work fills this gap by identifying photons as the agents
that establish thermal distributions, maintain characteristic energy
scales, and produce entropy through inelastic scattering.

\section{Steady state versus equilibrium}
\label{sec:steadystate}
Formal statistical mechanics defines thermal equilibrium as a
stationary state with detailed balance. It does not assert that
microscopic processes cease: a system in thermal equilibrium with
a radiation field continuously emits and absorbs photons, and no net
cooling occurs precisely because emission is balanced by absorption.
This is not in dispute. Our claim is narrower and more specific: the
common physical picture that the term ``thermal equilibrium'' conveys---
a state of stasis, of nothing happening---is wrong for any real system
at $E_c > 0$, and the formalism, by abstracting away the maintaining
mechanism, inadvertently reinforces that picture.

Any system of charged particles couples to the electromagnetic field
and continuously emits thermal radiation. Quantum electrodynamics
requires spontaneous photon emission from every excited state. An
infrared camera pointed at any object above 0~K directly images the
continuous outflow of photon energy. Emission is not optional and not
a small correction: a blackbody at 300~K radiates approximately
460~W~m$^{-2}$ to its environment at all times~\cite{jackson1999}.
When emission is not balanced by absorption, $E_c$ decreases: the
system cools. Constant $E_c$ therefore requires continuous compensating
input through photon absorption, internal energy generation, or both.
We emphasize the qualifier: it is not emission per se that causes
cooling, but \emph{net} radiative loss. In a closed cavity at
temperature $T$, walls emit and absorb continuously, and no net
cooling occurs. Cooling requires that the system be open, radiating
into a colder environment without compensating return flux. The
systems we are primarily concerned with are precisely such open systems,
or systems that would cool if their coupling to the environment were
severed.

The Boltzmann distribution
\begin{equation}
P(E) \propto e^{-E/E_c}, \qquad E_c \equiv k_BT
\label{eq:boltzmann}
\end{equation}
is stationary; its constituent excitations are in continuous exchange.
A useful analogy: a standing wave and the absence of oscillation are
both time-independent in appearance, but one involves continuous energy
flow and the other does not. The Boltzmann distribution of a real system
is like the standing wave---its form is steady, but it is sustained by
ongoing flux.
This leads to a physically natural classification. \textbf{Dynamic
steady state}: internal processes generate energy compensating radiative
losses. The system maintains constant $E_c$ while radiating to colder
surroundings---electrically heated blackbody cavities, stellar
photospheres (maintained by nuclear fusion), planetary surfaces
(maintained by absorbed solar radiation). \textbf{Passive steady state}:
the system exchanges photons with surroundings at matched $E_c$.
Absorption and emission balance with zero net energy flow---a sample
in thermal contact with a large bath, a gas in a thermostated enclosure.
The distinction is boundary-dependent: a sample in passive steady state
with its cryostat becomes part of a dynamically maintained system when
the cryostat is included. This boundary dependence reveals the
hierarchical structure of temperature maintenance discussed in
Sec.~\ref{sec:hierarchy}.
Replacing the static picture of ``equilibrium'' with the dynamic picture
of ``steady state'' is not semantic housekeeping. It immediately raises
the question: maintained by what? It directs attention to the energy
source sustaining the distribution. It clarifies why isolated systems
cool (they lose the mechanism of maintenance) and why temperature
cannot be assigned to a single isolated quantum (no ensemble, no steady
state is possible).

\section{Photon-mediated thermalization}
\label{sec:photons}
Any system of charged particles couples to the electromagnetic field.
Excited states decay through spontaneous emission with rates given by
Einstein $A$ coefficients~\cite{einstein1917,cohen1977}:
\begin{equation}
A_{21} = \frac{\omega^3|\langle 2|\mathbf{d}|1\rangle|^2}
{3\pi\epsilon_0\hbar c^3},
\end{equation}
where $\omega = (E_2-E_1)/\hbar$ and $\mathbf{d}$ is the electric
dipole operator. More generally, accelerating charges radiate with
power~\cite{jackson1999}
\begin{equation}
P = \frac{e^2 a^2}{6\pi\epsilon_0 c^3},
\end{equation}
where $a$ is the acceleration. Thermal motion inherently involves
charge acceleration; emission is mandatory for any charged system
with $E_c > 0$.
Einstein's detailed-balance analysis forces the radiation field to
follow the Planck distribution when matter follows the Boltzmann
distribution~\cite{milonni1994,scully1997}. The physical consequence
is immediate: a system at $E_c$ that emits to surroundings at lower
$E_c$---or to vacuum---loses energy net. Excited-state populations
decline. Without compensating absorption or internal generation,
$E_c$ relaxes downward. Constant $E_c$ requires that the system be
embedded in a radiation field of matched spectral density, or that
internal processes generate energy at the rate of radiative loss.

\section{The Maxwell distribution and the billiard-ball objection}
\label{sec:maxwell}
A natural and important objection to the photonic picture is the
following. Boltzmann's $H$-theorem~\cite{boltzmann1872} shows that a
gas of particles interacting through purely mechanical collisions---no
electromagnetic coupling, no photons---relaxes to the Maxwell-Boltzmann
velocity distribution
\begin{equation}
f(v) \propto v^2 \exp\!\left(-\frac{mv^2}{2E_c}\right)
\label{eq:maxwell}
\end{equation}
from any initial condition. The derivation makes no reference to photons.
If mechanical collisions alone can produce and maintain the thermal
distribution, in what sense is photon exchange necessary?
This objection is correct and important, and it must be answered carefully.
The $H$-theorem establishes that mechanical collisions drive a gas
\emph{toward} the Maxwell distribution Eq.~\eqref{eq:maxwell} for a
given total kinetic energy. It does not establish what determines or
maintains that total kinetic energy over time.
Consider an isolated gas of perfectly hard, electrically neutral spheres
in a perfectly insulating container. The $H$-theorem applies: the
velocity distribution relaxes to Maxwell form with $E_c$ determined
by the initial total kinetic energy. Once there, it stays---because
the gas is perfectly isolated. This is a mathematically consistent
idealization.
Now consider a real gas. Its molecules are composed of charged particles:
electrons and nuclei. They radiate. The Maxwell distribution is
established by collisions, yes---but $E_c$ is simultaneously being
eroded by radiative emission. In the absence of photon absorption from
the surroundings, $E_c$ decreases: the gas cools. The Maxwell
\emph{shape} is maintained at each instant by collisions, but the
\emph{scale} $E_c$ of that distribution is set and maintained by the
radiation field with which the gas is in exchange.
Collisions establish the thermal form of the distribution for the
current $E_c$. In open systems, the long-time value of $E_c$ is
controlled by energy exchange with the environment, including radiative
losses and gains. Both mechanisms are necessary for a complete
description of a real gas at approximately constant temperature: one
governs the distributional shape, the other governs the boundary
condition that stabilizes the energy scale.

\subsection{Internal redistribution versus boundary maintenance}
A further objection might invoke non-radiative mechanisms---phonon-phonon
scattering in solids, electron-phonon coupling, spin exchange---as
examples of thermalization proceeding without photon exchange. We accept
that these mechanisms efficiently redistribute energy internally and
are not in dispute. Our claim is different: it concerns the
\emph{boundary condition} that sets and stabilizes $E_c$ over time
against environmental losses, not the internal redistribution of
energy at fixed $E_c$.

Any surface at finite $E_c$ emits thermal radiation into its
surroundings regardless of which internal mechanism redistributes
energy within. Place a solid whose phonons redistribute energy
efficiently into an environment colder than $E_c$~---~or into a
radiation-free vacuum~---~and the solid will cool by emitting
into that environment. The phonons will continue to redistribute
energy internally, maintaining the Maxwell or Bose-Einstein form
of the internal distribution at each instant, but $E_c$ itself
will drift toward the environmental temperature scale. It is the
electromagnetic coupling at the boundary that sets the equilibrium
value of $E_c$, not the efficiency of internal redistribution.
Conversely, a material whose internal redistribution is slow
but whose boundary is in contact with a radiation field at
$E_c$ will thermalize toward that $E_c$ on the timescale set by
the boundary coupling, not the internal one.

This is precisely why the infinite-reservoir hierarchy of
Sec.~\ref{sec:hierarchy} has the structure it does: each level sets
the boundary condition for the level within. In the macroscopic
hierarchy relevant to laboratory and planetary systems, electromagnetic
energy exchange provides a ubiquitous boundary channel through which
temperature is stabilized across scales. The $H$-theorem and the boundary-exchange
mechanism are not in conflict: they operate at different levels,
one governing the distributional form given $E_c$, the other governing
the value of $E_c$ itself through the energy balance at system
boundaries.

\section{The mean blackbody photon energy: the 2.701 factor}
\label{sec:dynamic}
We derive the mean energy per photon in the three-dimensional Planck
distribution. This quantity characterizes the typical energy scale of
photons exchanged in any radiatively coupled system at temperature $T$,
and determines the spectral range over which photon exchange must occur
to sustain the full Boltzmann distribution of excited states.

The photon number density at frequency $\nu$
is~\cite{planck1901,pathria2011}:
\begin{equation}
n(\nu) = \frac{8\pi\nu^2}{c^3}\frac{1}{e^{h\nu/E_c}-1}.
\end{equation}
The average photon energy is:
\begin{equation}
\langle h\nu\rangle = \frac{\int_0^\infty h\nu\,n(\nu)\,d\nu}
{\int_0^\infty n(\nu)\,d\nu}
= E_c\,\frac{\int_0^\infty x^3/(e^x-1)\,dx}
{\int_0^\infty x^2/(e^x-1)\,dx},
\end{equation}
where $x = h\nu/E_c$. These are standard Bose-Einstein
integrals~\cite{pathria2011}:
\begin{align}
\int_0^\infty \frac{x^3}{e^x-1}\,dx &= \Gamma(4)\,\zeta(4)
= \frac{\pi^4}{15}, \\
\int_0^\infty \frac{x^2}{e^x-1}\,dx &= \Gamma(3)\,\zeta(3)
= 2\zeta(3),
\end{align}
where $\zeta(3) = 1.20206\ldots$ is Ap\'{e}ry's constant. Therefore:
\begin{equation}
\boxed{\langle h\nu\rangle = \frac{\pi^4}{30\,\zeta(3)}\,E_c
\approx 2.701\,E_c.}
\label{eq:2701}
\end{equation}
The mean photon energy exceeds $E_c$ because the Planck spectrum
carries a substantial high-frequency tail. While occupation numbers
fall exponentially as $n(\nu) \propto e^{-h\nu/E_c}$, energy per photon
rises linearly with frequency. Their product, the spectral energy
density $u(\nu) = h\nu\cdot n(\nu)$, peaks near
$h\nu_{\rm peak} \approx 2.82\,E_c$ (Wien's displacement
law~\cite{wien1896}).

The significance of Eq.~\eqref{eq:2701} is twofold. First, it
establishes that photons doing the work of maintaining a full Planck
spectrum---populating the high-frequency tail that excites the upper
states of matter---carry on average $2.701\,E_c$, not merely $E_c$.
Replacing only the mean thermal energy $\langle E\rangle \sim E_c$ per
degree of freedom is insufficient; the full spectral distribution, with
its high-frequency content, must be sustained. Second, and directly
observable, the factor 2.701 sets the mean photon energy scale in any
blackbody source, from the CMB at 2.7~K to stellar photospheres,
and governs the radiative coupling between systems at different
temperatures.

Photons also drive entropy production through inelastic scattering.
A single high-energy photon excites a system, which subsequently
decays through multiple channels emitting several lower-energy photons:
\begin{equation}
h\nu_{\rm in} = \sum_{i=1}^n h\nu_i, \qquad n > 1.
\end{equation}
Energy is conserved; photon number increases. The number of accessible
microstates scales as $W \sim (E_0/E_c)^{N-1}$ for $N$ photons sharing
energy $E_0$~\cite{reif1965,pathria2011}, so one ultraviolet photon
at 10~eV produces roughly ten infrared photons at 1~eV, increasing
entropy from $\sim 0$ to $\sim 21\,k_B$ while conserving energy
exactly. The time-reversed process---multiple photons spontaneously
combining into one---is allowed by conservation laws but has
vanishingly small probability without engineered nonlinear
conditions~\cite{boyd2008}. This asymmetry defines the thermodynamic
arrow of time: photon multiplication is spontaneous; photon
combination is not~\cite{lebowitz1993}.

\section{The infinite reservoir hierarchy}
\label{sec:hierarchy}
The cornerstone of the canonical ensemble is the thermal reservoir:
a system whose $E_c$ remains exactly fixed while exchanging arbitrary
energy with smaller systems~\cite{callen1985,fermi1956}. This idealization
is mathematically precise and underlies the success of the canonical
ensemble---it is the controlled thermodynamic limit in which $E_c$
changes negligibly on experimental timescales. We do not challenge the
mathematical legitimacy of this idealization. What we identify is the
physical basis on which it rests, and what it abstracts away.

For a reservoir's $E_c$ to remain fixed while giving energy to something
else, it must simultaneously receive compensating energy from somewhere
else. A reservoir that only gives energy and receives nothing cools.
An \emph{infinite} reservoir would need to supply arbitrary energy
indefinitely without replenishment. No such object exists. The infinite
reservoir is therefore not a physical entity. It is an effective
description of a system large enough that its $E_c$ changes negligibly
on the timescale of the experiment, and whose own temperature is
maintained by energy exchange at a yet larger scale.

Real thermal baths form a natural hierarchy: a cryogenic sample sits
in passive steady state with its cryostat, which requires continuous
electrical power to hold its 4~K against the 300~K laboratory, which
is maintained by HVAC systems, which balance against a planetary surface
maintained by solar flux, which is itself sustained by nuclear fusion
in a stellar interior. Each level appears as an effective infinite
reservoir to the levels within it because the timescale for significant
$E_c$ change at that level vastly exceeds experimental timescales at
the level below. The approximation is excellent and well-justified; its
basis is temporal scale separation, not ontological infinitude.

This perspective also clarifies that constant temperature does not
require an infinite reservoir. What it requires is a source of energy
flux that replaces losses faster than the experimental timescale.
In ordinary macroscopic systems---a blackbody cavity maintained by
electrical heating, the solar photosphere maintained by fusion, a
planetary surface maintained by absorbed solar radiation---electromagnetic
radiation is a ubiquitous and often dominant channel for that flux.
These are finite physical systems maintaining constant $E_c$ through
ongoing energy exchange, not by virtue of infinite heat capacity.

\section{Dimensionless entropy and microstate bookkeeping}
\label{sec:entropy}
In statistical mechanics, entropy is defined by
\begin{equation}
S = k_B \ln W,
\label{eq:boltzmann_entropy}
\end{equation}
where $W$ is the number of accessible microstates consistent with
the macroscopic constraints. The quantity with direct statistical
meaning is the \emph{dimensionless} entropy
\begin{equation}
\mathcal{S} \equiv \frac{S}{k_B} = \ln W,
\label{eq:dimensionless_entropy}
\end{equation}
which coincides with information-theoretic entropy in natural
units~\cite{jaynes1957a,shannon1948}. The factor $k_B$ converts this
dimensionless count into SI units (J/K) and, through the relation
$E_c = k_B T$, serves as the physical bridge connecting microstate
counts to macroscopic observables such as temperature and heat.
In that sense, $\mathcal{S} = \ln W$ carries the full combinatorial
content of thermodynamic entropy; $k_B$ converts it into the units
of the historical temperature scale without adding new statistical
information.

In the steady-state perspective of this paper, entropy production
is naturally associated with the growth of the accessible microstate
volume as energy is redistributed into a larger number of excitations.
This is precisely what occurs in photon multiplication: a high-energy
photon is converted into many lower-energy photons, increasing the
number of ways to distribute the same total energy among available
modes and thereby increasing $W$ and $\mathcal{S}$. The microstate
counting underlying $\mathcal{S} = \ln W$ depends on the appropriate
statistics, but the steady-state mechanism---continuous exchange
maintaining $E_c$ and enabling redistribution---applies regardless
of statistics.

It is important to distinguish \emph{energy levels} from
\emph{microstates}. Energy levels label eigenvalues of a Hamiltonian;
microstates label the underlying configurations---occupation patterns
of many modes, positions and momenta in phase space---consistent with
macroscopic constraints. Many distinct microstates can share the same
total energy, and the degeneracy generally grows rapidly with system
size. Therefore $\mathcal{S} = \ln W$ is not in general an integer
and should not be identified with an excitation index or energy
level number.

\section{Criteria for genuine temperature}
\label{sec:criteria}
Not all exponential distributions $P(E) \propto e^{-E/E_c}$ represent
physically realized temperatures. We establish criteria distinguishing
systems in genuine thermal steady states from systems where an
exponential fit is merely a formal statistical description.
For a \textbf{dynamic steady state}, genuine temperature requires a
continuous and observable energy-loss mechanism (Stefan-Boltzmann
emission), an identifiable internal energy source compensating those
losses, and spectral consistency: emitted radiation approximately
following the Planck distribution at the claimed $E_c$.
For a \textbf{passive steady state}, it requires an identifiable
thermal bath with well-characterized $E_c$, observable or inferable
photon exchange between system and bath, and a measured Boltzmann
distribution of internal-state populations verifiable through
spectroscopy.
The counterexample is instructive. A single hydrogen atom in vacuum
fails both tests. An excited state decays through spontaneous emission
with no mechanism to repopulate it. There is no ensemble of degrees
of freedom from which a statistical distribution can emerge. Temperature
is a collective property of systems undergoing continuous photon
exchange~\cite{goldstein2006,popescu2006}. Individual quanta have
energy but not temperature. This is not a measurement limitation; it
reflects the fundamentally collective and dynamic nature of the
steady-state distribution.

\section{Connection to classical thermodynamics}
\label{sec:connection}
Classical thermodynamics achieves its predictive success because the
exponential form Eq.~\eqref{eq:boltzmann} follows from statistical
requirements---additivity of energy, multiplicativity of
probability---that do not depend on what physical processes create
the distribution~\cite{jaynes1957a}. The theory correctly captures
the mathematical structure of stationary distributions independently
of mechanism. This is analogous to continuum fluid mechanics:
it correctly predicts bulk behavior without tracking molecules, because
the constitutive equations follow from symmetry and conservation
independently of molecular dynamics. Molecular dynamics explains
\emph{why} the continuum equations hold and identifies their limits.
Both levels are valid and serve different purposes.
The mechanistic level becomes necessary for non-equilibrium systems,
where fluctuation theorems~\cite{jarzynski1997,crooks1999,seifert2012}
and stochastic thermodynamics require explicit attention to system-bath
interactions; for small systems with few degrees of freedom where
fluctuations dominate and individual quantum transition rates become
relevant~\cite{esposito2009,campisi2011}; and when evaluating whether
an exponential distribution in an unusual system represents a genuine
temperature or a formal fit. In all these cases the criteria of
Sec.~\ref{sec:criteria} provide a framework that purely phenomenological
thermodynamics cannot.

\section{Summary}
\label{sec:conclusion}
We have argued that temperature in ordinary matter is a dynamically
maintained steady state, not a static equilibrium. The principal
results are:
(1) Any real open system at $E_c > 0$ continuously exchanges energy
with its environment through electromagnetic radiation. When net
emission exceeds absorption, $E_c$ decreases toward the environmental
temperature scale. ``Thermal equilibrium,'' far from implying stasis,
is physically a steady state maintained by ongoing energy exchange,
with electromagnetic radiation serving as a ubiquitous and often
dominant channel in open macroscopic systems. Formal statistical
mechanics correctly defines equilibrium as a stationary state with
detailed balance; what the present work identifies is the energy
exchange process that physically sustains this condition in real
open systems.

(2) The mean photon energy in the Planck distribution is
$\langle h\nu\rangle \approx 2.701\,E_c$
(Eq.~\eqref{eq:2701}), derived from the ratio of standard Bose-Einstein
integrals. This characterizes the typical energy scale of photon
exchange in any radiatively coupled system and quantifies why
maintaining the high-frequency tail of the Planck spectrum requires
photons carrying well above $E_c$.

(3) The Maxwell velocity distribution of a gas is established by
mechanical collisions, while in open systems the long-time value of
$E_c$ is controlled by energy exchange with the environment.
Collisions govern the distributional form; boundary exchange governs
the energy scale. Internal redistribution mechanisms such as
phonon scattering operate within the system but do not replace the
boundary condition: any surface at finite $E_c$ emits into its
environment, and the equilibrium value of $E_c$ is set by the balance
at that boundary, with electromagnetic radiation providing a ubiquitous
and dominant channel.

(4) Every finite thermal reservoir is maintained by energy exchange
at a larger scale. The classical infinite reservoir is the
large-capacity limit of a physical hierarchy, valid when the
surrounding bath's $E_c$ changes negligibly on experimental timescales.
Its physical basis is temporal scale separation; it is a controlled
idealization, not a fundamental physical entity.

(5) The dimensionless entropy $\mathcal{S} = \ln W$ carries the full
statistical content of thermodynamic entropy; $k_B$ converts to SI
units and links microstate counts to macroscopic temperature and energy
scales. Entropy production corresponds to growth of the accessible
microstate volume, made vivid by photon multiplication in inelastic
scattering.

These results do not modify thermodynamics but provide its mechanistic
foundation in quantum electrodynamics. Classical theory correctly
characterizes stationary distributions; the photonic mechanism
identifies what physically realizes and sustains them.

\begin{acknowledgments}
This work was supported by the U.S.\ Department of Energy, Office of
Science, Basic Energy Sciences, Materials Sciences and Engineering
Division. Ames National Laboratory is operated for the U.S.\
Department of Energy by Iowa State University under Contract
No.\ DE-AC02-07CH11358.
\end{acknowledgments}

\bibliographystyle{apsrev4-2}

\end{document}